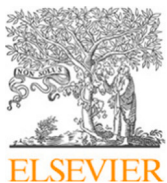
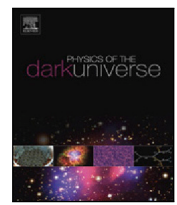

Full length article

# Quantum isotropic Universe in RQM analogy: The cosmological horizon

Gabriele Barca [a,b], Luisa Boglioni [a,*], Giovanni Montani [c,a]

[a] *Department of Physics, "Sapienza" University of Rome, P.le Aldo Moro 5, Rome, 00185, Italy*
[b] *INFN, Sezione di Roma 1, P.le A. Moro 2, Rome, 00185, Italy*
[c] *ENEA, Fusion and Nuclear Safety Department, C. R. Frascati, Via E. Fermi 45, Frascati (RM), 00044, Italy*

ABSTRACT

We investigate the quantum dynamics of the isotropic Universe in the presence of a free massless scalar field, playing the role of a physical clock. The Hilbert space is constructed via a direct analogy between the Wheeler-DeWitt equation in the minisuperspace and a relativistic scalar one in physical space. In particular, we show how the introduction of a "turning point" in the Universe evolution allows to overcome an intrinsic ambiguity in representing the expanding and collapsing Universe. In this way, the positive and negative frequencies are simply identified with time reversed states. The main subject of the present analysis is the construction of a horizon operator, whose quantum behavior is investigated when Polymer Quantum Mechanics is implemented to describe the asymptotic evolution near the initial singularity. The reason of this choice is motivated by the intrinsic spreading of localized wavepackets when the polymer dispersion relation governs the quantum dynamics. The evidence that the mean value of the quantum horizon operator follows its semiclassical behavior (corrected for polymerization) is a clear indication that a concept of causality can be restored also in the quantum cosmological picture.

## 1. Introduction

The implementation of quantum physics to the isotropic Universe evolution is affected by a significant number of shortcomings [1,2]. However, many significant results have been achieved during the years, both in the metric and Loop formulations [3–7]. The possibility for a bouncing cosmology [8–11], the emergence of a classical dynamics when the anisotropies are small enough [12–15] and the possibility for a Born–Oppenheimer separation for the system evolution [16] are to be regarded as concrete physical insights on the possible quantum origin of our Universe.

An important role is played in cosmology by the basic concept of the particle horizon, which fixes the scale of causality from the initial singularity up to a given instant of the Universe age [17]. This concept, being related to the motion of photons across the Universe, loses a clear meaning in the quantum picture. Indeed, the Universe wavefunction seems to describe the evolution of the Universe as a whole, completely disregarding any notion of quantum causality.

In this paper, the question of recovering a notion of causality for a quantum Universe is faced in the context of a direct analogy between the minisuperspace of the quantum isotropic Universe, in the presence of a free massless scalar field used as a relational time, and the Relativistic quantum dynamics of a scalar particle in physical spacetime (for similar discussions see [18,19]).

In this approach, an ambiguity regarding the idea of an expanding and collapsing Universe is discovered, and immediately solved with the introduction of a "turning point" for the wavepackets describing the evolution of the Universe. The quantum dynamics of the considered system is analyzed in the framework of Polymer Quantum Mechanics [20], when the evolution of initially localized states is naturally associated with a spreading behavior towards the initial singularity. This scenario is identified as the most appropriate in which we can test the predictivity of our proposal for a quantum horizon operator, developed by implementing the canonical prescription on its classical expression.

The most relevant result provided by the present analysis is the evidence that the mean value of the quantum horizon well reproduces the classical behavior of the causality scale. In other words, the proposed approach to the restoration of a notion of causality in the quantum primordial Universe appears to be a predictive procedure. Therefore, this could be easily extendable to more general contexts in order to understand if the Universe wavefunction is still able to establish a causality relation between different parts of space, and which implications this property could have on the physical nature of the emerging quantum Universe.

## 2. Standard cosmology

### 2.1. WDW equation

The Standard Cosmological Model (SCM) is based on the Friedmann-Lemaître-Robertson-Walker (FLRW) model, a non-stationary homogeneous and isotropic three-geometry. The ADM line element for this model is

$$ds^2 = -N(t)^2 dt^2 + a(t)^2 dl^2_{RW}, \quad (1)$$

where $a(t)$ is the scale factor of the Universe, $N(t)$ is the lapse function of the ADM decomposition [21–23], and the term $dl^2_{RW}$ denotes the spatial line element of a three-space, which for the flat case is given by

$$dl^2_{RW} = \delta_{ij} dx^i dx^j. \quad (2)$$

Let us consider a Universe described by the metric (1) and filled with a massless scalar field $\Phi$. From the formula of the ADM decomposition, we can find the expression for the extrinsic curvature of the minisuperspace[1] as:

$$K_{ij} = \frac{1}{2N}\left(\partial_0 h_{ij} - D_i N_j - D_j N_i\right), \quad (3)$$

where $N^i$ represents the shift vector. The Ricci scalar reads as

$$R = {}^3R + K_{ij} K^{ij} - K^2 - 2\nabla_\mu(\eta^\nu \nabla_\nu \eta^\mu - \eta^\mu K), \quad (4)$$

where ${}^3R$ is the three-dimensional one, and $\eta^\nu$ is a time-like vector. The action results to be[2]

$$\begin{aligned}S &= \int_{\mathbb{R}\times\Sigma} dx^0 d^3x \mathcal{L} \\ &= V_0 \int_{\mathbb{R}} dx^0 \sqrt{-g}\left(\frac{1}{16\pi}R - \frac{1}{2}g^{\mu\nu}\partial_\mu\Phi\partial_\nu\Phi\right) \\ &= V_0 \int_{\mathbb{R}} dx^0 \left(-\frac{3}{8\pi}\frac{\dot{a}(t)^2 a(t)}{N(t)} + \frac{1}{2}\frac{\dot{\Phi}^2}{N(t)}a(t)^3\right).\end{aligned} \quad (5)$$

In order to move from the Lagrangian formalism to the Hamiltonian one, we calculate the conjugate momenta:

$$\Pi_a = \frac{\partial \mathcal{L}}{\partial \dot{a}} = -\frac{3}{4\pi}\frac{a(t)\dot{a}(t)}{N(t)}, \quad (6)$$

$$\Pi_\Phi = \frac{\partial \mathcal{L}}{\partial \dot{\Phi}} = \frac{\dot{\Phi}a(t)^3}{N(t)}, \quad (7)$$

$$\Pi_N = \frac{\partial \mathcal{L}}{\partial \dot{N}} = 0. \quad (8)$$

Note how the momentum conjugate to $N$ vanishes. This fact has a central role in the analysis of the system. Indeed, the Hamiltonian density takes the form

$$H = N\left(-\frac{2\pi}{3}\frac{\Pi_a^2}{a} + \frac{1}{2}\frac{\Pi_\Phi^2}{a^3}\right). \quad (9)$$

If we operate the canonical transformations to the new variables

$$\Phi \to \Phi' = A\Phi = \sqrt{\frac{3}{4\pi}}\Phi, \quad (10)$$

$$a(t) = e^\alpha, \quad -\infty < \alpha < +\infty, \quad (11)$$

the new Hamiltonian density is[3]

$$H = N\frac{2\pi}{3}e^{-3\alpha}\left(-\Pi_\alpha^2 + \Pi_\Phi^2\right). \quad (12)$$

---
[1] See Cap. 10 in [1].
[2] From the homogeneity hypothesis, we are able to integrate out the spatial dependence of the three-geometry from the action integral. In the case of a flat Universe, we choose for convenience to integrate over a fiduciary volume $V_0$ of the same value that in the further calculation we set to 1.
[3] After the transformation we drop the prime on the scalar field to thin notation.

Now, the Poisson brackets for the variable $N$ vanishes:

$$\{H, \Pi_N\} = 0 = \frac{\delta H}{\delta N} = \mathcal{H}. \quad (13)$$

We find that the so-called super-Hamiltonian $\mathcal{H}$[4] coincides with a secondary constraints, so we deal with a constrained Hamiltonian system.

Let us notice that if we calculate the classical evolution of the variables through the Hamilton equations, we obtain

$$\dot{\Phi} = \frac{\partial H}{\partial \Pi_\Phi} = \frac{4\pi}{3}Ne^{-3\alpha}\Pi_\Phi, \quad \dot{\Pi}_\Phi = -\frac{\partial H}{\partial \Phi} = 0, \quad (14)$$

$$\dot{\alpha} = \frac{\partial H}{\partial \Pi_\alpha} = -\frac{4\pi}{3}Ne^{-3\alpha}\Pi_\alpha, \quad \dot{\Pi}_\alpha = -\frac{\partial H}{\partial \alpha} = 0. \quad (15)$$

In order to solve the Problem of Time (POT) [24–27], we consider the massless scalar field as a time coordinate and so the classical equation of motion becomes

$$\frac{d\alpha}{d\Phi} = \frac{\dot{\alpha}}{\dot{\Phi}} = -\frac{\Pi_\alpha}{\Pi_\Phi}. \quad (16)$$

Now we move on to quantization, using the canonical paradigm. We can promote the variables of our system to operators as

$$\alpha(t) \to \hat{\alpha} \equiv \alpha, \quad \Pi_\alpha \to \hat{\Pi}_\alpha \equiv -i\frac{\delta}{\delta\alpha}, \quad (17)$$

$$N(t) \to \hat{N} \equiv N, \quad \Pi \to \hat{\Pi} \equiv -i\frac{\delta}{\delta N}, \quad (18)$$

$$\Phi(t) \to \hat{\Phi} \equiv \Phi, \quad \Pi_\Phi \to \hat{\Pi}_\Phi \equiv -i\frac{\delta}{\delta\Phi}. \quad (19)$$

We can introduce the concept of a wavefunction of the Universe

$$\Psi = \Psi(\alpha, \Phi), \quad (20)$$

and we can use the Dirac prescription to implement the classical constraint as a quantum operator that annihilates the physical states. The quantum equation for the dynamics then is,

$$\hat{H}\Psi(\alpha, \Phi) = \frac{2\pi}{3}Ne^{-3\alpha}\left(-\frac{\partial^2}{\partial\Phi^2} + \frac{\partial^2}{\partial\alpha^2}\right)\Psi(\alpha, \Phi) = 0. \quad (21)$$

This is the so-called Wheeler–DeWitt equation [28], for the FLRW flat Universe.

### 2.2. The wavefunction of the universe

The expression above represents the quantum dynamical equation for the gravitation field in the presence of a scalar field. Let us notice that it is very similar to a Klein–Gordon equation. Therefore, it can be treated as a Klein–Gordon equation where the time coordinate is represented by the massless scalar field $\Phi$, while the position coordinates correspond to the logarithmic scale factor $\alpha$. Therefore, the solutions are easy to derive through Relativistic Quantum Mechanics.

Firstly, we consider the plane-wave solution:

$$\psi = Ae^{i(k\alpha - \omega\Phi)}. \quad (22)$$

We can find the dispersion relation, as

$$\omega = \pm|k|, \quad (23)$$

that leads to two different solutions: the positive-frequency solution and the negative-frequency one. We want to construct a normalized, Gaussian wavepacket that represents an expanding Universe. By looking at the classical evolution, the expansion or contraction of the Universe depends on the sign of $\dot{\alpha}$. From Eq. (15), that still contains synchronous time $t$, the evolution of the Universe depends only on the sign of $\Pi_\alpha$. Therefore, in the classical dynamics with time $t$, there are four different scenarios for our system represented in the following table:

---
[4] See Cap.11 in [17].

|  | $\Pi_\alpha > 0$ | $\Pi_\alpha < 0$ |
|---|---|---|
| $\Pi_\Phi > 0$ | Collapse, $\Phi$ grows | Expansion, $\Phi$ grows |
| $\Pi_\Phi < 0$ | Collapse, $\Phi$ decreases | Expansion, $\Phi$ decreases |

These four different physical situations disappear when we decide to use the scalar field $\Phi$ as time coordinate. As we can see from Eq. (16), the evolution of the Universe depends only on the ratio of the two momenta, so we can fix one of them and consider only the sign of the other one. For example, by considering a negative sign for the momentum conjugate to the scale factor, the negative sign of $\Pi_\Phi$ corresponds to a collapse, while the positive one to an expanding system. On the other hand, the positive sign of $\Pi_\alpha$ corresponds to the opposite situations for the momentum of the scalar field. So there are two identical copies of the Universe with opposite signs for the two momenta.

Now, when dealing with highly localized wavepackets, it is possible to relate the classical values of the momenta (i.e., $\Pi_\alpha$ and $\Pi_\Phi$) to the corresponding quantum eigenvalues (i.e., $k$ and $\omega$, respectively), in agreement with the Ehrenfest theorem. Hence, in the quantum realm, changing the sign of $\Pi_\Phi$ means taking $-\omega$ instead of $\omega$, that in the particle physics representation replaces "particles" with the corresponding "antiparticles". We can say that this interpretation leads to two copies of the Universe that are completely equivalent, but where particles with antiparticles are switched; when talking about wavepackets, the entire spectrum of $k$ then gives redundant information, allowing us to restrict the available Hilbert space. Therefore, the expression of the wavefunction of an expanding Universe results to be

$$\Psi(\alpha, \Phi) = \int_{-\infty}^{0} dk \frac{1}{\sqrt{2\pi}\sqrt{2|k|}} G(k) e^{ik(\alpha - \Phi)}, \tag{24}$$

where

$$G(k) = \sqrt{\frac{2}{\text{Erfc}(\bar{k}/\sigma)\sigma\sqrt{\pi}}} e^{-\frac{(k-\bar{k})^2}{2\sigma^2}} \tag{25}$$

is the normalized Gaussian profile.[5] So, according to the "Correspondence Principle", the classical ambiguity in the sign of $\Pi_\alpha$ would corresponds to the set up of two different wave packets, having positive or negative mean value of the operator $\hat{\Pi}_\alpha$, respectively. Here, we prefer to remove this restriction, by reconnecting the expanding and collapsing branches, considering a curved, closed Universe. In such a system Eq. (16) changes. The two flat-case trajectories, which correspond to the two bisectors in the $(\Phi - \alpha)$ plane, are no longer separate situations. Instead, the Universe expands, reaches the so-called "turning point", and then starts to re-collapse. Therefore, in this case the choice of the sign of $\Pi_\Phi$ does not select the trajectory of the Universe, but just the direction of time. In the particle representation we can say that a "particle", i.e. the positive-frequency solution $e^{-ik\Phi}$, follows the trajectory with the standard direction of time, while an "anti-particle", the "negative-frequency" solution $e^{ik\Phi}$, follows the inverse trajectory backwards in time. However, since the aim of the present analysis is to study the behavior of the horizon operator, the inclusion of the curvature has non-trivial implications on the locality of the theory. We use here a simplified, but reliable scheme: we reconnect the expanding and collapsing branches by a boundary condition, ensuring the vanishing behavior of the Universe wave function for a given (maximum) value of the variable $\alpha$.[6] In order to do that, we impose a cut-off in $\alpha$ for the wavepacket. We redefine $\Psi$ by imposing a "barrier" at an arbitrary $\alpha_{max}$, requiring that

$$\Psi(\alpha_{max}, \Phi) = 0. \tag{26}$$

---

[5] The peak $\bar{k}$ of the Gaussian and the variance $\sigma$ are fixed by the initial conditions on the wavefunction at a given initial time $\Phi_0$ (Erfc is the conjugate error function).

[6] The addressed scenario implies a weak violation of the theory unitarity, in the sense that it is asymptotically restored toward the singularity (where we will estimate the horizon mean value).

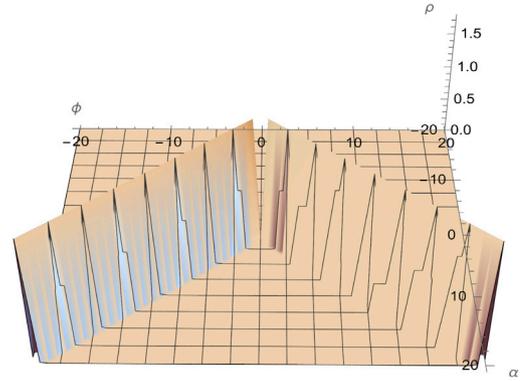

**Fig. 1.** Probability density of the wavepacket $\Psi(\alpha, \Phi)$, with $\bar{k} = -10$ and $\sigma = 3$.

This is achieved by performing the transformation

$$\Psi(\alpha, \Phi) \to \Psi((\alpha - \alpha_{max}), \Phi) - \Psi(-(\alpha - \alpha_{max}), \Phi) \tag{27}$$

and inserting a Heaviside step function $\Theta$. This way the condition (26) is satisfied and the new wave packet $\Psi(\alpha, \Phi)$ becomes

$$\Psi(\alpha, \Phi) = \int_{-\infty}^{0} dk \frac{iG(k)}{\sqrt{\pi|k|}} \sin(k(\alpha_{max} - \alpha))\Theta(\alpha_{max} - \alpha)e^{-ik\Phi}. \tag{28}$$

The behavior of our wavepacket, represented in Fig. 1, is then given by its Klein–Gordon-like probability density:

$$\rho_\Phi = i(\Psi^* \partial_\Phi \Psi - \Psi \partial_\Phi \Psi^*). \tag{29}$$

The wavepacket follows the expected trajectory, describing an expanding Universe that, after a "turning point", starts to recollapse. The wavepacket remains picked throughout all the evolution, up to the singularity at $\Phi \to \pm\infty$. So, with the expression (28), the semiclassical description is valid for the entire evolution.

## 3. Modified cosmology

The final aim of this work is to restore the concept of horizon, and thus causality, in the primordial Universe. However, moving towards the singularity, a semiclassical description is not sufficient anymore because quantum gravitational effects become relevant. In order to account for that, we will construct a wavepacket within the framework of the Polymer representation.

### 3.1. Polymer representation

In quantum cosmological theories, a discrete structure of spacetime seems to be a fundamental feature of the Early Universe [1,29,30]. In order to describe mathematically such a physical cut-off in the generalized coordinate at microscopical scales, an alternative representation of quantum mechanics is necessary. This is the aim of Polymer Quantum Mechanics (PQM), [20].

The main assumption of this non-regular representation of Quantum Mechanics is that one of the variables (for example the position) of a two-dimensional phase space is discretized. As a consequence, its conjugate momentum can be promoted as an operator only after a regularization. This leads to the introduction of a graph, i.e. a lattice structure on the space.

From a mathematical point of view, we firstly define a basis for the Hilbert space $\mathcal{H}_{poly}$ as normalized abstract kets $|\mu\rangle$ labeled by a real number $\mu$. A generic cylindrical state can be written as

$$|\psi\rangle = \sum_{i=1}^{N} a_i |\mu_i\rangle, \tag{30}$$

where the kets $|\mu_i\rangle$ are labeled by a finite collection of numbers $\mu_i \in \mathbb{R}$ with $i = 1, 2, \ldots, N$. Now, we can introduce two basic operators on this Hilbert space, the symmetric label operator $\hat{\epsilon}$ and the unitary shift operator $\hat{s}(\lambda)$:

$$\hat{\epsilon}|\mu\rangle := \mu|\mu\rangle, \tag{31}$$

$$\hat{s}(\lambda)|\mu\rangle := |\mu + \lambda\rangle; \tag{32}$$

note that $\hat{s}$ is discontinuous in the parameter $\lambda \in \mathbb{R}$. In a standard system with a discrete configuration coordinate $q$ and its conjugate momentum $p$, we can denote the states as

$$\psi(p) = \langle p|\psi\rangle, \tag{33}$$

where

$$\psi_\mu(p) = \langle p|\psi\rangle = e^{i\frac{\mu p}{\hbar}}. \tag{34}$$

In this representation, the position operator in the momentum polarization acts as

$$\hat{q}\psi_\mu(p) = -i\hbar\frac{\partial}{\partial p}\psi_\mu(p) = \mu e^{i\frac{\mu p}{\hbar}} = \mu\psi_\mu(p), \tag{35}$$

and therefore we identify $\hat{q}$ with the abstract label operator $\hat{\epsilon}$. On the other hand, the shift operator becomes the exponential operator $\hat{S}(\lambda)$:

$$\hat{S}(\lambda)\psi_\mu(p) = e^{i\frac{\lambda p}{\hbar}}e^{i\frac{\mu p}{\hbar}} = e^{i\frac{(\mu+\lambda)p}{\hbar}} = \psi_{\mu+\lambda}(p) \tag{36}$$

and it is discontinuous. Since a discontinuous operator cannot be generated by a hermitian one by exponentiation, it is impossible to define $\hat{p}$ rigorously.

As a regularization, we introduce a regular graph $\gamma_{\mu_0}$ with constant step $\mu_0$:

$$\gamma_{\mu_0} = \{q \in \mathbb{R} : q = n\mu_0, \forall n \in \mathbb{Z}\}. \tag{37}$$

Now a regulated operator $\hat{p}_{\mu_0}$ that depends on the scale $\mu_0$ can be defined as

$$\hat{p}_{\mu_0}|\mu_n\rangle = \frac{\hbar}{2i\mu_0}(|\mu_{n+1}\rangle - |\mu_{n-1}\rangle). \tag{38}$$

Actually, for $p \ll \hbar/\mu_0$, one obtains

$$\hat{p} \sim \frac{\hbar}{\mu_0}\sin(\frac{\mu_0 p}{\hbar}). \tag{39}$$

This implies that the eigenvalues of the momentum operator are limited from above by $\hbar/\mu_0$.

### 3.2. Polymer cosmology

Now, we want to implement the polymer paradigm to our model of Universe. We define the action of the momentum conjugate to $\alpha$ as

$$\hat{\Pi}_\alpha \sim \frac{\sin\mu\Pi_\alpha}{\mu}, \tag{40}$$

and therefore the Wheeler-DeWitt equation becomes

$$\left(-\frac{\sin[2](\mu\Pi_\alpha)}{\mu^2} + \Pi_\Phi^2\right)\psi(\Pi_\alpha, \Phi) = 0, \tag{41}$$

(we set $\hbar = 1$ and drop the subscript 0 on $\mu$ to thin notation). Note that the wavefunction of the Universe depends on $\Pi_\alpha$ and $\Phi$ because the polymer representation is more straightforward in the $p$-polarization. The solution to this equation is

$$\psi(\Pi_\alpha, \Phi) = \delta(k_\alpha - \tilde{k}_\alpha)e^{-ik\Phi}, \tag{42}$$

where $\tilde{k}_\alpha = \frac{\arcsin(k\mu)}{\mu}$. The dispersion relation then is

$$k = \pm\left|\frac{\sin\mu k_\alpha}{\mu}\right|. \tag{43}$$

This new dispersion relation is not linear, so a spreading wavepacket is expected.

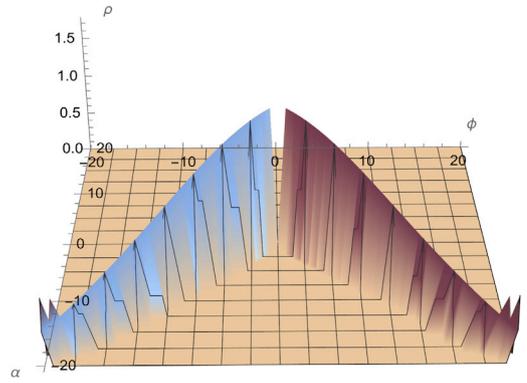

**Fig. 2.** Probability density of the wavepacket $\Psi(\alpha, \Phi)$ in polymer representation, with $\bar{k} = -10$, $\sigma = 3$ and $\mu = 1/30$.

We perform a Fourier transformation in order to have its expression in coordinate representation for both the variables, and we impose the cut-off, as before. The new wavepacket, setting $\alpha_{max} = 0$ for simplicity, becomes

$$\Psi(\alpha, \Phi) = \int_{-\frac{1}{\mu}}^{0} \frac{idkG(k)}{\sqrt[4]{1-\mu^2k^2}\sqrt{\pi k}}\sin(\tilde{k}_\alpha\alpha)e^{-ik\Phi}\Theta(-\alpha), \tag{44}$$

where the Gaussian profile in this case is defined as

$$G(k) = \sqrt{\frac{2}{\mathrm{Erfc}(\frac{1+\bar{k}\mu}{\sigma\mu}) - \mathrm{Erfc}(\frac{\bar{k}}{\sigma})}}\frac{e^{-\frac{(k-\bar{k})^2}{2\sigma^2}}}{\sqrt[4]{\pi\sigma^2}} \tag{45}$$

in order to be normalized in the interval $-1/\mu \leq k \leq 0$. Fig. 2 shows the behavior of this new wavepacket.

The wavepacket follows the classical trajectory as in the standard case, but here there is an important difference. Approaching the singularity at smaller values of $\Phi$, the variance in the probability density becomes larger, showing the spreading of the wavepacket during evolution. Therefore, a semiclassical description is not sufficient anymore, since at some point the variance will become of the same order as the expectation value and a fully quantum treatment will be necessary.

### 4. Quantum concept of the horizon

The cosmological horizon is defined as the maximal casual distance that a photon has traveled from the Big Bang up to today. This concept, called particle horizon, corresponds to every particle signal that has reached the observer between the time of the Big Bang ($t = 0$) and the time $t$. Considering particles traveling radially at light speed towards the observer, the horizon is calculated from the condition for the propagation of a wave front, i.e.

$$ds^2 = 0, \tag{46}$$

that leads to the expression

$$d_H(t) = a(t)\int_0^t \frac{dt'}{a(t')}. \tag{47}$$

However, on a quantum level, the concept of trajectory and the idea of a point particle disappear. Therefore, the definition of cosmological horizon described above, strictly connected with the idea of the path traveled by photons, is not well-defined. In order to restore this concept, the following approach can be pursued: we can find a quantum counterpart for the horizon, promoting it to operator, and evaluate the viability of this idea comparing its expectation value with the semiclassical solution.

The expression of the horizon in our system, after the ADM decomposition and some mathematical passages, is

$$d_H = e^\alpha \int_{-\infty}^\alpha d\tilde{\alpha}\frac{Ne^{-\tilde{\alpha}}}{\mathrm{H}(\tilde{\alpha}, \Phi)}, \tag{48}$$

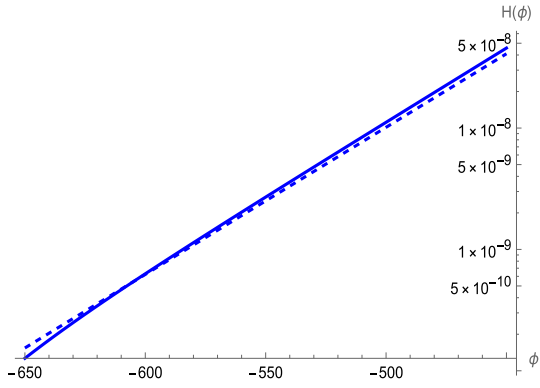

**Fig. 3.** Logarithmic plot of mean value of the horizon (continuous line) compared with the semiclassical value (dashed line), with $\mu = 1/30$, $\bar{k} = -10$, $\sigma = 3$ and $\Pi_\Phi = 11$. In this plot the variable $\Phi$ is rescaled in $\Phi/100$ in order to facilitate the numerical integration near the singularity.

where $\mathrm{H}(\tilde{\alpha}, \Phi)$ is the Hubble parameter, that for a Universe filled only with a massless scalar field results to be

$$\mathrm{H}(\alpha, \Phi) = \dot{\alpha} = \frac{4\pi}{3} N e^{-3\alpha} \Pi_\Phi. \tag{49}$$

The horizon then becomes

$$d_H(\alpha, \Pi_\Phi) = \frac{3}{4\pi} e^\alpha \int_{-\infty}^{\alpha} d\tilde{\alpha} \, \frac{e^{2\tilde{\alpha}}}{\Pi_\Phi}. \tag{50}$$

Now, in order to quantize the horizon, we promote the momentum conjugate to the scalar field to a quantum operator in the canonical way:

$$\Pi_\Phi \to \hat{\Pi}_\Phi = -i \frac{\partial}{\partial \Phi}. \tag{51}$$

As it is claimed in [31,32], a common representation of the inverse momentum operator in Hilbert space is

$$\frac{1}{P} \Psi(x) = i \int_{-\infty}^{x} \Psi(x') dx' \tag{52}$$

where $\Psi(x)$ must satisfy the condition[7]

$$\int_{-\infty}^{+\infty} \Psi(x) dx = 0 \tag{53}$$

to belong to the domain of the operator $\hat{P}^{-1}$. According to that, we obtain the "cosmological horizon" operator as

$$\hat{d}_H = i \frac{3}{8\pi} e^{3\alpha} \int_{-\infty}^{\Phi} d\Phi'. \tag{54}$$

We evaluate the expectation value of this quantum operator on the wavefunction of the Universe, as defined in (44).

The semiclassical (sc) evolution is reached considering the effective polymer Hamiltonian derived from Eq. (41), that leads to a modified expression for the Hubble parameter:

$$\mathrm{H}^{sc}(\alpha, \Phi) = \dot{\alpha} = \frac{4\pi}{3} N e^{3\alpha} \Pi_\Phi \sqrt{1 - \mu^2 \Pi_\Phi^2}. \tag{55}$$

Therefore the semiclassical expression for the horizon results to be

$$d_H^{sc}(\Phi) = \frac{3}{8\pi} e^{3\Phi} \frac{1}{\Pi_\Phi \sqrt{1 - \mu^2 \Pi_\Phi^2}}. \tag{56}$$

In Fig. 3 we show the quantum expectation value of the horizon operator[8] compared with its corresponding semiclassical trajectory.

---

[7] This condition is satisfied by our wave packet.
[8] The boundary condition in the $\alpha$-space induces a small violation of the unitarity, vanishing toward the singularity.

Note how the expectation value of the quantum operator reproduces the semiclassical behavior with good agreement. The quantum notion of causality reflects its classical counterpart. This is the most interesting result of the analysis, which allowed to restore the concept of causality also in the first instants of the Universe.

## 5. Concluding remarks

The analysis above had the scope to set up a reliable interpretative framework for analyzing the introduction of a quantum operator, which could be able to provide information on the causality of different space regions of a quantum Universe.

The starting point was the construction of suitable (initially localized) states for the isotropic Universe in the presence of a free massless scalar field, used as physical clock for the quantum dynamics (according to the relational idea of time [33]). The Hilbert space which provides a probability distribution for such states has been set up in close analogy to the standard procedure of Relativistic Quantum Mechanics [34]. In fact, the Wheeler-DeWitt equation outlines a structure isomorphic to that of a $1 + 1$ Klein–Gordon theory. In this respect, in order to provide the correct Feynman interpretation of the positive and negative frequency solutions as associated to the two possible direction of the time arrow [35], a turning point has been introduced for the macroscopic Universe via a suitable boundary condition on the physical states.

Then, the horizon operator has been inferred from its classical expression, when stated in terms of physical quantities that are well-identifiable in the quantum regime too. Since the standard metric approach provided non-spreading localized states, the reliability of using the conjectured causal operator has been evaluated on states constructed through the implementation of Polymer Quantum Mechanics [20,29]. The polymerization procedure induced a non-linear dispersion relation for the eigenfunctions of the system, and a spreading behavior is observed towards the singularity (that is still present in the adopted variable [36,37]).

The very interesting feature of the proposed horizon operator is that, even on spreading states that give an intrinsically quantum representations of the dynamics, the expectation value as a function of time overlaps the classical dynamics, once corrected for polymerization [38]. This result, together with the existing link between Polymer Quantum Mechanics and Loop Quantum Cosmology [6,11,39], leads us to infer that our notion of horizon is an intriguing perspective for future more general implementations of the same idea. In this sense, we can say that, even in a quantum isotropic Universe, it is still possible to recover spatial regions which have less probability than others to be in causal contact, i.e. able to reach a kind of "equilibrium", fine tuning the nature of the fluctuating physical quantities.

## CRediT authorship contribution statement

**Gabriele Barca:** Formal analysis, Resources, Writing – review & editing. **Luisa Boglioni:** Formal analysis, Methodology, Resources, Writing – original draft, Writing – review & editing. **Giovanni Montani:** Conceptualization, Supervision.

## Declaration of competing interest

The authors declare that they have no known competing financial interests or personal relationships that could have appeared to influence the work reported in this paper.

## Data availability

No data was used for the research described in the article.